# *Квантовая Жизнь Реваза Ильича Джибути*

## Роман Я. Кезерашвили

31 августа 2014 исполнилось бы **80 лет** одному из известных грузинских и советских физиков, Лауреату Государственной Премии СССР, доктору физико-математических наук, профессору **Ревазу Ильичу Джибути.**

Реваз Ильич Джибути был одним из учеников патриарха ядерной физики в Грузии академика Вагана Ивановича Мамасахлисова. В конце 1971 года я был зачислен в аспирантуру Тбилисского Государственного Университета и был последним аспирантом Вагана Ивановича, а Реваз Ильич Джибути был со руководителем. Через полгода Ваган Иванович скончался, и я стал первым официальным аспирантом Реваза Ильича. В своих воспоминаниях я хотел бы отразить квантовую жизнь **Реваза Ильича Джибути** с позиции его ученика, сотрудника возглавляемого им отдела Теоретической Ядерной Физики в Институте Физики АН Грузинской ССР, его соавтора и просто физика, который имел возможность работать с ним до конца его дней.

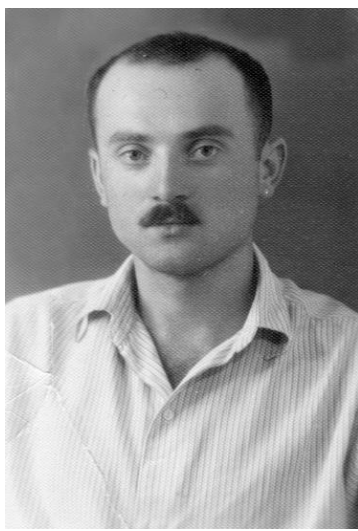

Реваз Джибути аспирант кафедры теоретической физики

В конце пятидесятых, - в начале шестидесятых годов большое количество проведенных экспериментальных и теоретических исследований структуры атомных ядер указывали на большую роль пространственных корреляций в ядрах. В частности, эксперименты по взаимодействию γ - квантов высоких энергий и по захвату π-мезонов атомными ядрами свидетельствовали о том, что у ядерных волновых функций существуют фурье-компоненты с высокими импульсами. Иными словами, ядерные нуклоны, обладающие большими импульсами, должны быть сильно коррелированы и находиться на более близком расстоянии, чем среднее расстояние между

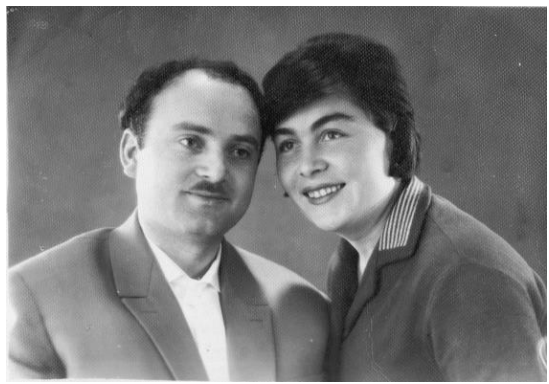

Реваз Джибути и его супруга Нели Чхиквишвили в год защиты кандидатской диссертации

нуклонами в ядре. Однако, теоретические модели господствующие в то время, предполагали движение нуклонов в среднем поле, созданном ядерными нуклонами или наличие кластерных ассоциаций в атомном ядре. Именно в этот судьбоносный момент развития ядерной физики Реваз Джибути начал свою научную деятельность по исследованию парных и многочастичных корреляций нуклонов в ядрах и в ядерных реакциях на легких ядрах. Его первая научная статья, написанная в соавторстве с А.В. Тагвиашвили, посвященная исследованию фотодезинтеграции ядра $^4$He фотонами высоких энергий была опубликована в самом престижном в то время советском журнале ЖЭТФ, который был наиболее широко известен за рубежом. Эта работа положила основу циклу работ по фоторасщеплению легких ядер с выходом протонов, нейтронов и нейтрон-протонных пар, выполненных Ревазом Джибути, и которые стали основой его кандидатской диссертации. 7 сентября 1963 года Реваз Ильич успешно защитил кандидатскую диссертацию на тему: "Некоторые ядерные реакции и корреляция нуклонов в легких ядрах". Его научным руководителем был академик Ваган Иванович Мамасахлисов, а Ученым секретарем Совета в то время был Тенгиз Иванович Санадзе.

После защиты кандидатской диссертации научные исследования Джибути были

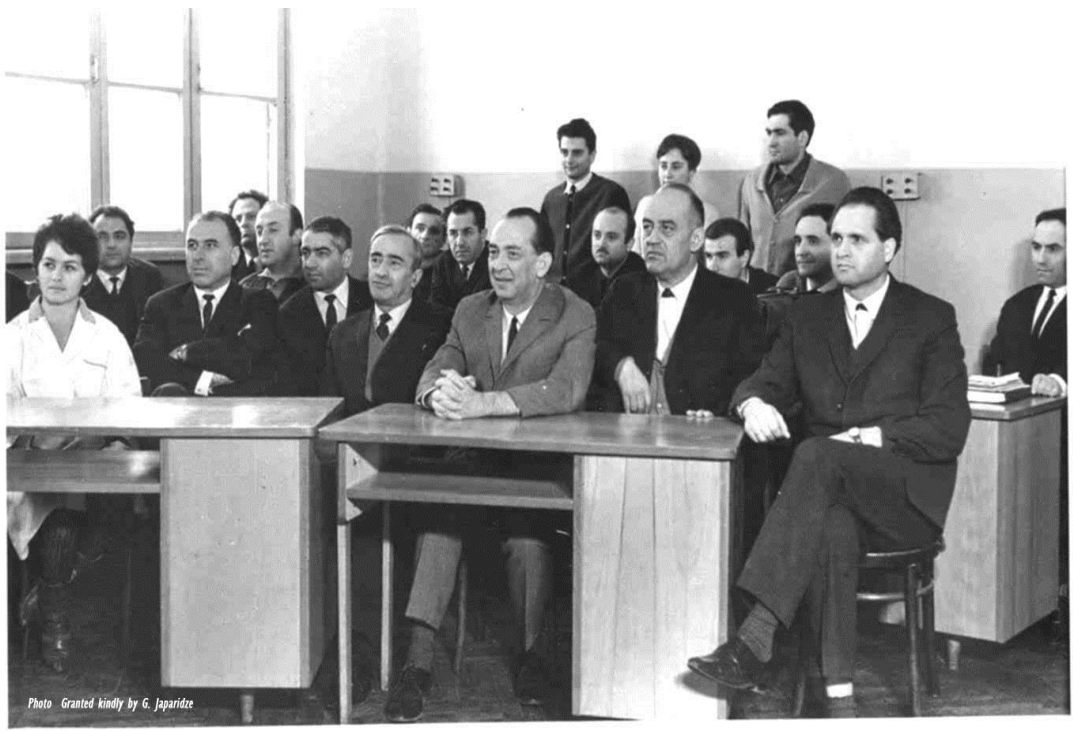

Семинар на кафедре теоретической физики ТГУ в начале шестидесятых. Семинар был основан Ваган Ивановичем Мамасахлисовым и проводился еженедельно по средам в 2 часа. Этот семинар был «Меккой» грузинских физиков-теоретиков.

Слева на право: Передний ряд: не опознана, Бикенти Вашакидзе, Ваган Мамасахлисов, Ника Полиектов Николадзе, Гиви Хуцишвили, Гурам Чилашвили; Второй ряд: Гиви Сурамлишвили, Вахтанг Тускиа, Нодар Цинцадзе, Сергей Матинян, Раваз Джибути, не опознан, Ираклий Мачабели, не опознан; Задний ряд: Тенгиз Мачарадзе, Гурам Долидзе, не опознан; Кетеван Костанашвили, не опознан; не опознан.

сосредоточены на продолжении изучения нуклонных корреляций в ядрах, в фотоядерных реакциях и в ядерных реакциях поглощения π-мезонов легкими ядрами. Интересно отметить, что пионерские исследования Джибути и Копалейшвили по выбиванию коррелированных пар нуклонов из ядер в реакциях с фотонами и пионами, опубликованные 1963-1964 годах, получили высокую признательность международной научной общественности и были отмечены в обзоре Джералда Брауна, который до кончины в 2013 был директором Института Теоретической Физики им. Янга в Нью-Йорке, основателем и редактором международных журналов Physics Letters и Physics Reports и редактором журнала Nuclear Physics.

В середине шестидесятых годов Реваз Джибути интенсивно продолжал работать по теории дезинтеграции легких ядер, рассматривая процессы выбивания нуклонов, дейтронов и α–частиц из ядер, изучал энергетические спектры и угловые распределения нуклонов в процессах фоторасщепления. К этому периоду его деятельности относится его идея получения эффективных ядерных потенциалов из реалистических нуклон-нуклоных потенциалов.

Одна из основных задач ядерной физики заключается в нахождении и знании нуклон-нуклонного взаимодействия, которое обычно строится на основе нуклон-нуклонного и нуклон-ядерного рассеяния. Потенциал взаимодействия между двумя нуклонами вне атомного ядра принято называть реалистическим потенциалом, тогда как нукон-нуклонный потенциал внутри ядра называют эффективным потенциалом, подчеркивая тем самым то обстоятельство, что присутствие ядерной среды влияет на нукон-нуклонное взаимодействие в ядре. Отличие между этими потенциалами следует ожидать особенно сильным на малых межнуклонных расстояниях, где эффект выхода вне энергетическую поверхность, являющийся главной причиной этого отличия, становится особо важным. Именно эту важную проблему "атаковал" Реваз Джибути в середине шестидесятых и предложил рецепт получения эффективных потенциалов путем перехода к явно

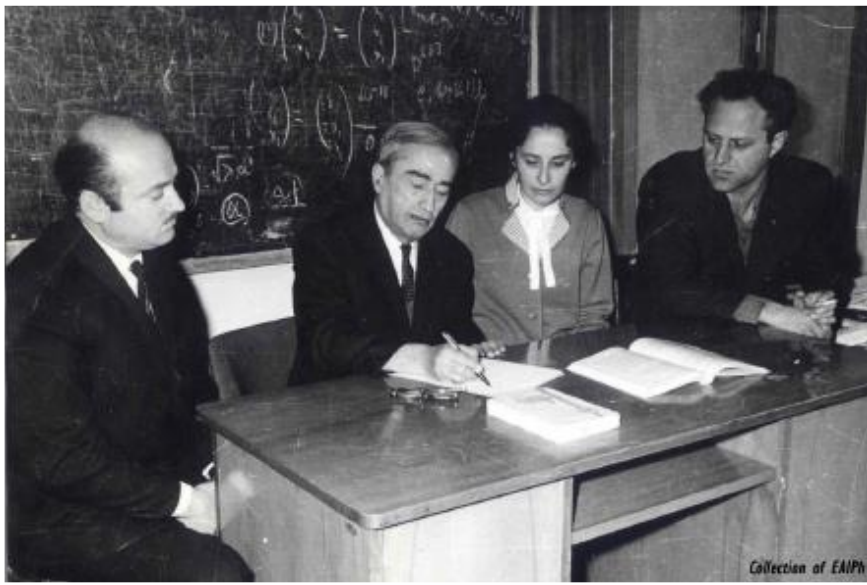

.И. Мамасахлисов со своими учениками. Слева на право: Р.И. Джибути, В.И. Мамасахлисов, Н.Б. Крупенникова, Т.С. Мачарадзе.

зависящему от скорости нелокальному потенциалу, сохраняя радиальную форму и параметры реалистического нуклон-нуклонного взаимодействия. Правило перехода от реалистического потенциала к эффективному являлось общим и не приводило к коллапсу ядерной материи. Интересно отметить, что предложенный эффективный нелокальный потенциал приводит к точному решению уравнения Бете-Голдстоуна для конечных ядер при любой форме реалистического потенциала. Предложенный эффективный, зависящий от скорости, потенциал интенсивно использовался в дальнейшем Ревазом Джибути и его коллабораторами для исследования интегральных сечений дипольного фотопоглощения, электромагнитных свойств легких ядер, расчета магнитных моментов легких атомных ядер, изучения обменных токов в ядерных реакциях и понимания важности эффективных сил в трехчастичных ядрах. На основе цикла работ связанных с зависящим от скорости потенциалом Нина Борисовна Крупенникова в 1970 защитила кандидатскую диссертацию на тему: "Зависящие от скорости нуклон-нуклонные потенциалы и электромагнитные свойства легких ядер". Нина работала в Институте Физики и была соискателем. Ее диссертация была первой кандидатской диссертацией выполненной под научным руководством Р.И. Джибути.

Надо отметить, что Реваз Ильич предложил новый подход к исследованию электромагнитных взаимодействий с ядрами в котором, в отличии от стандартного подхода, электромагнитные и ядерные эффекты заранее не разделялись, что дает возможность через электромагнитные свойства ядер получать непосредственную информацию о специфических свойствах ядерных сил. Работы по фотоядерным реакциям, и по изучению электромагнитных свойств легких ядер на основе идеи предложенного эффективного зависящего от скорости потенциала составили костяк докторской диссертации Р.И. Джибути: "Эффективные нуклон-нуклонные потенциалы и электромагнитные взаимодействия в ядрах". Защита диссертации состоялась в ноябре 1971 года и официальные оппоненты были профессор В. В. Бабиков из ОИЯИ, профессор И.Ш. Вашакидзе и профессор С.Г. Матинян. В ноябре прошлого 2013 года я встретился с профессором С.Г. Матиняном в Северной Каролине, США в Central University of North Carolina, Durham и рассказал ему что в октябре 2014 мы устраиваем в Тбилиси Симпозиум, посвященный 80-летию Реваза Джибути. Сергей Гайкович задумался и потом сказал: "Резо был прекрасный физик. Еще в середине шестидесятых он предложил рассматривать эффективные, зависящие от скорости взаимодействия. Сегодня подходы на основе эффективных лагранжианов составляют одно из главных направлений в ядерной физике низких энергий, в частности, в низкоэнергетической квантовой хромодинамике".

Исследование малочастичных систем всегда привлекало внимание теоретиков и экспериментаторов в течение долгого времени и продолжает оставаться актуальным в наши дни. Конец шестидесятых - начало семидесятых годов ознаменовался возросшим интересом к исследованию малочастичных систем в связи с успехом методов уравнений Фаддеева и Фаддеева-Якубовского и метода гиперсферических функций. Соответственно, в начале семидисятых годов научные исследования Реваза Джибути были сфокусированы в двух направлениях. С одной стороны, это исследование короткодействующих динамических многочастичных корреляций в ядерной материи и в реакциях рассеяния электронов высоких энергий на ядрах, с другой стороны, изучение структуры малочастичных ядер в рамках метода гиперсферических функций. Это было то время когда я стал аспирантом Реваза Ильича. Реваз Ильич поставил задачу решения уравнения Бете-Фаддеева с целью исследования вклада трехчастичных короткодействующих динамических корреляций в энергию связи ядерной материи. В дальнейшем эти исследования переросли в изучение четырехчастичных корреляций в ядерной материи и нами были обобщены уравнения Фаддеева-Якубовского для четырех тел на случай четырех нуклонов в ядерной материи. Результаты этих разработок и исследования короткодействующих корреляций в ядерных реакциях позже были подытожены в монографии Реваза Ильича *"Динамические корреляции нуклонов в атомном ядре"*. Это было замечательное время, когда каждое утро мы с Ревазом Ильичом обсуждали результаты, полученные в ночных расчетах на ЕВМ М-220М (задачи нашего отдела обычно мы приносили в вычислительный центр в конце рабочего дня и Вова Томчинский называл этот поход "пошли в ночное", так как наши задачи были времяемкими и считались ночью). Обычно в это же время Вова Томчинский приносил результаты его расчетов по формфакторам и энергии связи ядер $^3$Не и $^3$Н методом гиперсферических функций и Реваз Ильич с Вовой и Ниной продолжал анализировать результаты ночных расчетов. В середине семидесятых годов в комнате где, находились Реваз Ильич, Нина Крупенникова, Вова Томчинский и я, всегда царила творческая атмосфера, наполненная дискуссиями и это придавало стимул работать не наблюдая часов. Я вспоминаю, как однажды после жарких обсуждений поздно вечером мы вчетвером вышли из института и

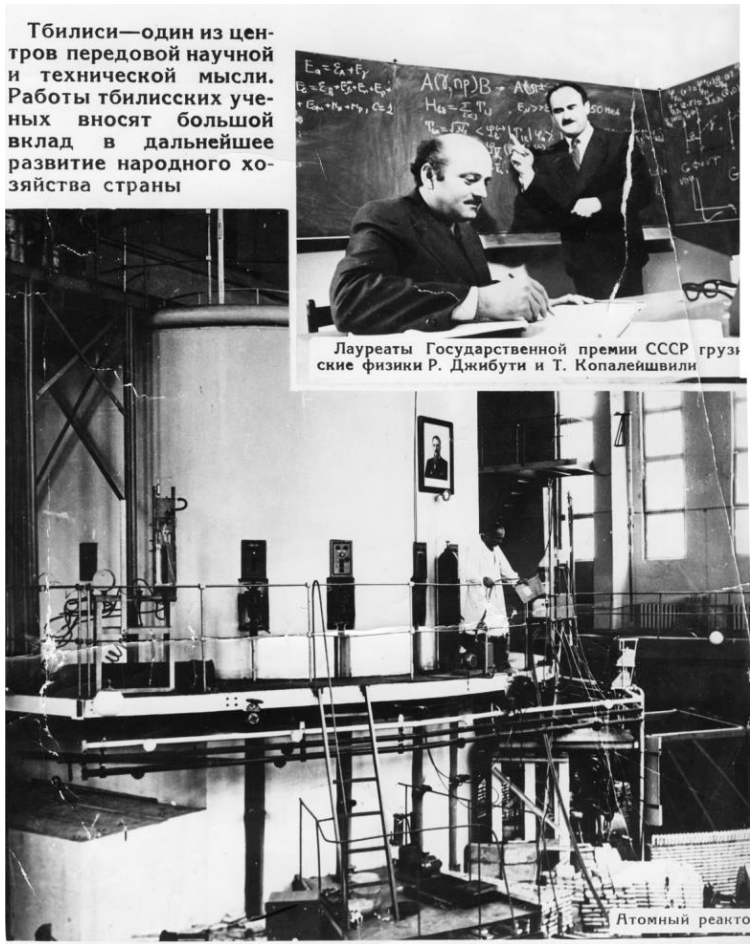

Реваз Ильич сказал: поспорили мы сегодня хорошо, но мы все равно остаемся сильно связанным четырехчастичным кластером!

В семидесятых годах Реваз Ильич и я опубликовали большой цикл работ по упругому и квазиупругому рассеянию электронов высоких энергий на легких ядрах. В частности, на основе рассмотрения многочастичных динамических корреляций был предсказан минимум форм-фактора упругого рассеяния электронов на ядре $^4$Не, который спустя три года был подтвержден в эксперименте на Стенфордском ускорителе. Наши исследования по квазиупругому рассеянию электронов высоких энергий на легких ядрах, в частности, предсказание наличия минимума в сечении рассеяния в процессе (e,e'p) на ядре $^4$Не обусловленного многочастичными короткодействующими динамическими корреляциями, стимулировали постановку экспериментов на Харьковском линейном электронном ускорителе, и к дальнейшей коллаборации с коллегами из Харьковского Физико-Технического Института. Я был следующий ученик Реваза Ильича, который в 1976 году представил к защите диссертацию на соискание ученной степени кандидата физико-математических наук на тему: "Многочастичные короткодействующие динамические корреляции в ядрах" [1].

В 1977 году грузинским физикам Ревазу Джибути и Теймуразу Копалейшвили вместе с группой физиков-экспериментаторов из Москвы и Харькова была присуждена Государственная Премия СССР за пионерские исследования по фотоядерным реакциям. Теоретические работы Джибути и Копалейшвили являлись основополагающими в теоретическом предсказании и мотивации экспериментальных исследований в этой области. В связи с Премией, мне хочется вспомнить банкет у Реваза Ильича дома, где столы ломились от явст, приготовленных верной саратницей Реваза Ильича калбатони Нели, где собрались сотрудники, друзья и ученики Реваза Ильича, где Зураб Саралидзе был тамадой и где столько прекрасных слов было сказано в адрес Резо Джибути - человека и ученого. А разве можно забыть калбатони Нелис фирменное и знаменитое "нигвзиани толма", вкус которого я и сейчас помню!

В 1979 году Элевтер Луарсабович Андроникашвили организонал в институте отдел ядерной физики руководителем которого был назначен Реваз Ильич. Исследования в отделе были сфокусированы в двух направлениях: исследования по теоретической ядерной физике и экспериментальные исследованаия с использованием нейтронных пучков атомного реактора Института Физики. Эксперименталые исследования по нейтронной ядерной оптике были новым направлением развиваемым в институте и Реваз Ильич Джибути с энтузиазмом руководил этим направлением Эксперименты проводились на реакторе группой Мурмана Цулая. Эти исследования получили широкую известность и позволили развить коллаборацию с Объединенным Институтом Ядерных Исследований в Дубне и другими научными центрами.

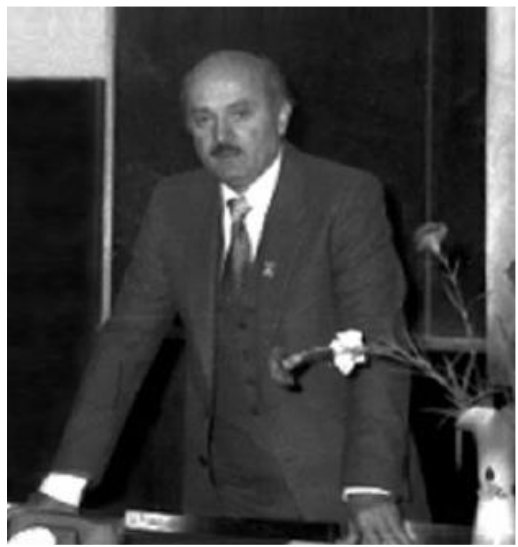

10 октября 1985 г. Р.И. Джибути открывает совещание, посвященное исследованиям по нейтронной ядерной оптике

Под руководством Р.И. Джибути, начиная с 1972 года и до конца его дней в Грузии начались изучение легких ядер в рамках метода гиперсферических функций. На первом этапе эти исследования были сфокусированы на формулировке основных уравнений метода К-гармоник для трехчастичных кластерных конфигураций, вычислении зарядовых формфакторов ядер $^6$Li, $^9$Be, и $^{12}$C в трехчастичной кластерной модели в координатном представлении. В дальнейшем главный акцент исследований был сосредоточен на развитие математического аппарата метода гиперсферических функций. В частности, были получены коэффициенты преобразования в гиперсферическом базисе для четырех частиц с неравными массами, установлена связь пятичастичных коэффициентов Рейнал-Реваи с четырехчастичными и двухчастичными коэффициентами Талми-Мошинского. В конце семидесятых публикуются пионерские работы по гиперсферическому базису в импульсном представлении и формулировке гиперрадиальных уравнений для трех и четырех тел в импульсном представлении. К этому времени относится развитие метода гиперсферических функций для описания нескольких частиц в непретывном спектре. В рамках метода гиперсферических функции разрабатываются методы исследования ядерных реакций с несколькими частицами в конечном состоянии. Это новое слово в изучении ядерных реакций развала с учетом взаимодействия всех продуктов реакции в конечном состоянии. Результаты этих исследований стали основой кандидатских диссертаций учеников Реваза Ильича - Владимира Томчинского, представившего диссертацию на тему: "Проблема четырех, пяти тел и ядерные реакции с тремя частицами в конечном состоянии", которую он защитил в 1977 году и Нодара Шубитидзе представившего диссертацию на тему: "Проблема четырех, пяти тел в гиперсферическом базисе", защищенную четырьмя годами позже, в 1981 году. В 1984 году Р.И. Джибути совместно с Н.Б. Крупенниковой публикует монографию: *"Метод гиперсферических функций в квантовой механике нескольких тел"*[3], где приводятся и обсуждаются история и новейшие математические разработки метода.

Реваз Джибути делал много докладов, в том числе приглашенных, на всесоюзных и международных конференциях и совещаниях. Каждый год профессор Джибути и его ученики участвовали в Ежегодном Совещании по Ядерной Спектроскопии и Структуре Атомного Ядра, которое проводилось в разных городах Советского Союза, в том числе и в Тбилиси. В 1975 году эта конференция проводилась в Ленинграде и Реваз Ильич, Нина Крупенникова, Вова Томчинский и я были участниками этой конференции. В это же время нашему коллеге Тенгизу Мачарадзе делали операцию в Москве. Мы постоянно связывались по телефону и узнавали состояние здоровья Тенгиза. Операция по удалению большой опухоли в голове длилась много часов и, к сожалению, не увенчалась успехом. Когда мы

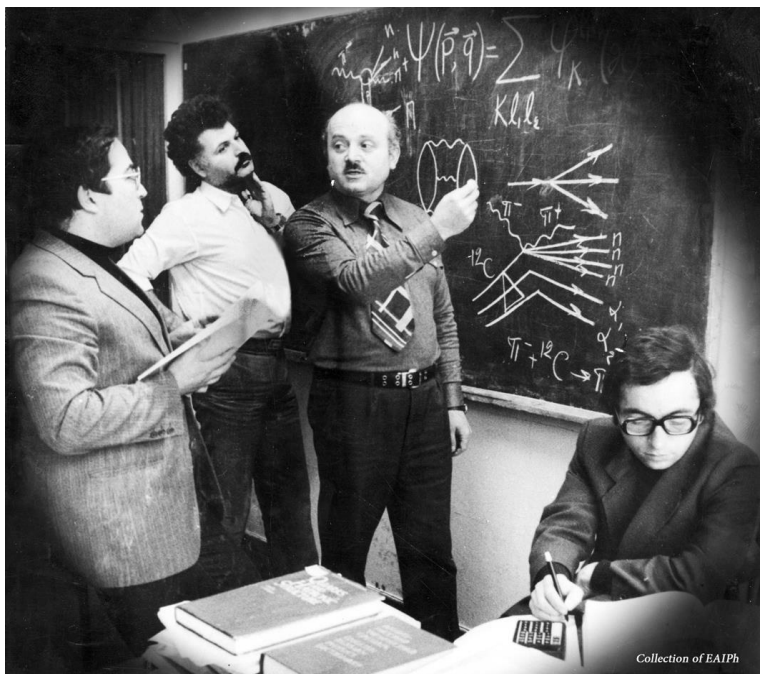

Реваз Джибути со своими учениками. Обсуждение реакции двойной перезарядки пионов на ядре $^{12}C$. Слева на право: Н.И Шубитидзе, Р.Я. Кезерашвили, Р.И. Джибути, Ш.М. Циклаури.

узнали о кончине Тенгиза, по настоянию Реваза Ильича мы немедленно вернулись в Тбилиси. Но хватит о печальном. Вспоминается такой курьезный случай, когда в 1978 году эта конференция состоялась в Баку. Из Тбилиси была вся группа Реваза Ильича и профессор И.Ш. Вашакидзе из Тбилиского университета. Когда мы приехали селиться в гостиницу там, как обычно, не было одноместных номеров и Реваз Ильич и Иван Шалвович попросили администратора поселить их вместе в двухместном номере. Но администратор ни в какую не соглашалась выполнить эту просьбу (увы, человеческой тупости нет предела!) и тогда Джибути и Вашакидзе обратились в оргкомитет конференции помочь им. Представитель оргкомитета пришел и стал объяснять администратору с сильным бакинским акцентом: "слюшай, адын профессор хочит спат с другим профессором, да. Слюшай, почему, да, ти нэ хочеш?" Джибути и Вашакидзе, и мы все вокруг покатывались со смеху.

Большой цикл работ Р.И. Джибути с сотрудниками был посвящен разработке методов и исследованию ядерных реакций с несколькими частицами в конечном состоянии в рамках метода гиперсферических функций. Была разработана теория "истинно" трех и черытехчастичного рассеяния, сформулированы вариационные принципы Хюльтона, Кона и Швингера для 3→3 и 4→ 4 рассеяния. Метод фазовых функции был обобщен для процессов с тремя и четырьмя частицами. В этот период Реваз Ильич предложил метод учета кулоновского взаимодействия в рамках гиперсферического базиса. Была осуществлена обширная программа детального изучения реакций полного развала трех и четырех нуклонных ядер вызванных фотонами, пионами и мюонами с учетом взаимодействия в конечном состоянии и, что особенно важно, начальные и конечные ядерные состояния описывались в рамках единого метода и находились с использованием одного и того же нуклон-нуклонного потенциала. Последнее обстоятельство обеспечивало ортогональность начального и конечного ядерных состояний и тем самым предотвращало ложный вклад в сечение процессов вызванных неортогональностью. Большое внимание было уделено изучению механизмов реакций. Были исследованы различные механизмы реакции двойной перезарядки пионов на ядрах и был предложен новый квази-$\alpha$-частичный механизм двойной перезарядки пионов на ядрах. Это было также то время, когда Р.И. Джибути предложил новую программу исследований малочастичных атомных и молекулярных систем методом гиперсферических гармоник. В частности, была предложена микроскопическая теория фотодезинтеграции атома гелия в гиперсферическом подходе. Результаты этих исследований составили основу диссертационных работ его учеников. В 1983 году Кетино Сигуа защищает диссертацию на тему: "Исследование некоторых многочастичных ядерных реакций методом гиперсферического базиса", а диссертация на тему "Микроскопический подход к исследованию трех и четырехчастичных реакций и многочастичного кулоновского рассеяния" была защищена Платоном Имнадзе в 1989 году. Диссертационная работа Демури Тедорадзе, ученика Реваза Ильича, который специализировался на изучении легчайших атомов, была посвящена исследованию атомных систем в использованием метода гиперсферических функций. Параллельно с использованием гиперсферического базиса в отделе Джибути проводились исследования структуры трех и четырехчастичных ядер с использованием осцилляторного базиса, начатые Тенгизом Мачарадзе. Эти исследования стали основой кандидатской диссертации Тенгиза Михелашвили на тему

"Исследование малонуклонных систем методом осцилляторного базиса" защищенной в 1980 году, научными руководителями которого были Реваз Джибути и Тенгиз Мачарадзе.

В начале восьмидесятых годов Реваз Ильич загорелся идеями гиперядерной физики. Гиперядро – это атомное ядро, в котором на ряду с нуклонами имеется по крайней мере один гиперон, - $\Lambda$– гиперон или $\Sigma$ – гиперон. В рамках метода гиперсферических функции под руководством Джибути были разработаны теоретические подходы расчета основных и возбужденных состояний трех, четырех и пятичастичных гиперядер состоящих из двух, трех и четырех нуклонов и одного гиперона или двух гиперонов. Эти подходы были обобщены на многочастичные гиперядра имеющие кластерную структуру. Были разработаны математические методы симметризации гиперсферических функций для описания гиперядер. В этот же период Реваз Ильич разрабатывает единый подход для описания связанных состояний и реакций распада гиперядер. Одна из интересных идей Реваза Ильича связана с "гибридным методом" в котором предлагается решение уравнений Фаддеева используя метод гиперсферических функций. Этот метод нашел свою реализацию при описании гиперядер. Результаты исследований по малочастичной гиперядерной физике составили основу диссертационных работ учеников Реваза Ильича. Шалва Циклаури защитил кандидатскую диссертацию на тему: "Исследование малочастичных гиперядерных систем используя метод гиперсферических функций" в 1986 году, а Лия Чачанидзе защитила кандидатскую диссертацию на тему: "Исследование четырех и пятичастичных гиперядер методом гиперсферических функций" в 1994 году, уже после кончины Реваза Ильича.

В одной из последних работ Р.И. Джибути использовал гиперсферический подход для описания трех кварковых структур и легких барионов и тем самым Реваз Ильич как бы завершил логическую цепь: кварки-ядра-атомы-молекулы. Итоги этой логической цепи подведены в монографии Р.И. Джибути: *"Метод гиперсферических функций в атомной и ядерной физике"* [6], написанной в соавторстве К.В. Шитиковой, которая была принята в печать при жизни Реваза Ильича, но опубликована 1993 году, после его кончины. Помимо основной научной деятельности Реваз Ильич Джибути проводил большую педагогическую работу. Несколько строк о профессоре Джибути как о преподавателе, педагоге и научном руководителе. Реваз Ильич читал мне курс теоретической ядерной физики, когда я был студентом четвертого и пятого курсов физического факультета Тбилисского Университета. Его лекции отличались невероятной ясностью,

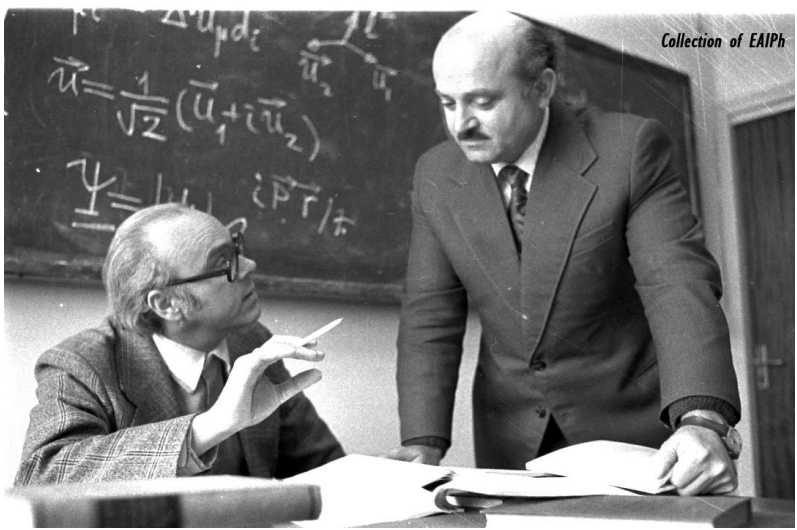

Одна из последних фотографий Р.И. Джибути. Г.А. Харадзе и Р.И. Джибути в Институте Физики

точностью и интересностью изложения, материал преподносился в таком ключе, что мотивировал студента критически осмысливать теоретические положения и предпосылки, что способствовало генерации идей и вопроса: А что еще здесь можно сделать? В принципе, именно после его курса я выбрал специализацию по теоретической ядерной физике. Какой научный руководитель был Реваз Ильич, по всей видимости можно судить хотя бы по одному показателю: под его руководством защитили кандидатские диссертации одиннадцать аспирантов и соискателей (в алфавитном порядке): П.М. Имнадзе, Р.Я. Кезерашвили, Н.Б. Крупенникова, Т.Я. Михелашвили, Н.М. Саллам, К.И. Сигуа, Д.К. Тедорадзе, В.Ю. Томчинский, Ш.Л. Циклаури, Л.Л. Чачанидзе, Н.И. Шубитидзе. Кроме этого под научным руководством Р.И. Джибути десятки студентов физического факультета ТГУ выполнили дипломные работы. В знак моего глубокого уважения и признания в день пятидесятилетия Реваза Ильича 31 августа 1984 года я подарил ему кубок с лаконичной надписью: "Дорогому Учителю от Ученика". Фото кубка, который по сей день уже тридцать лет хранится в семье Джибути, любезно предоставил мне младший сын Реваза Ильича Гоча. Как важный вклад Р.И. Джибути в педагогику мне хочется упомянуть учебники, написанные Ревазом Ильичом на грузинском и русском языках: *"ლექციები მრავალი ნაწილაკის კვანტურ მექანიკაში"*[2], *"Методы теоретической ядерной физики"* [4], *"თეორიული ბირთვული ფიზიკის მეთოდები"* [5] опубликованные в 1982, 1988 и 1991 годах.

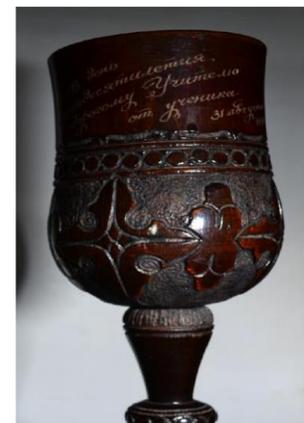

В день пятидесятилетия «Дорогому Учителю от Ученика»

Обычно живущие закрывают глаза усопшим и только некоторые усопшие могут открыть глаза живущим... Реваз Ильич Джибути ушел из жизни в полном расцвете творческих сил. За короткий период своей научной деятельности он выполнил обширную программу научных исследований и стал известным ученым, физиком-теоретиком мирового класса в области ядерной физики и исследования взаимодействий элементарных частиц с нуклонами и ядрами. Я не перестаю удивляться и восхищаться широте его научных интересов которые простирались от кварков до атомов и молекул. На грузинском языке никогда не говорят "умер" о людях ушедших из жизни, а говорят "гардаицвала", что в переводе на русский язык означает "переменился", "перевоплатился". Так и Реваз Ильич лишь перевоплатился и ушел из нашего ИЗМЕРЕНИЯ в другое, неведомое нам ПЯТОЕ ИЗМЕРЕНИЕ, а его мысли и дела живут в памяти его последователей.

Литература

1. **Р.И. Джибути**, Динамические корреляции нуклонов в атомном ядре. Мецниереба, Тбилиси, 1981, 144 с . (*In Russion*)

2. **რ. ჯიბუტი,** ლექციები მრავალი ნაწილაკის კვანტურ მექანიკაში. თბილისის უნივერსიტეტის გამომცემლობა, 1982, 203 გვ. (*In Georgian*).

3. **Р.И. Джибути,** Н.Б. Крупенникова, Метод гиперсферических функций в квантовой механике нескольких тел. Мецниереба, Тбилиси, 1984. 181с. (*In Russion*).